\documentclass[aip,graphicx,algorithm,algpseudocode]{revtex4-1}

\usepackage{amsmath,amssymb,xcolor}

\newcommand{\needcite}{[{\color{red} cite}]}
\newcommand{\DBWY}[1]{{\color{red}{DBWY: #1}}}

\newif\ifexporttikz
\newif\ifusetikz

\exporttikzfalse  % for importing tikz figures
\usetikzfalse     % for importing tikz figures without Tikz

\ifexporttikz
    \usepackage{tikz}
    \usepackage{pgfplots}
    \pgfplotsset{compat=newest}
    \usepgfplotslibrary{external}
    \usepgfplotslibrary{fillbetween}
    \usetikzlibrary{patterns, intersections}
\else
    \ifusetikz
      \usepackage{tikzexternal}
      \tikzsetexternalprefix{fig-export/}
    \else
      \usepackage{graphicx}
    \fi
\fi

\ifusetikz
  
  \newcommand{\importlocalfigure}[1]{%
    \figname{#1}
    \input{fig-src/#1.tex}
  }
\else
  \newcommand{\importlocalfigure}[1]{%
    \includegraphics{fig-export/#1.pdf}
  }
\fi

\usepackage{newfloat,algcompatible}
\usepackage{algpseudocode}
\DeclareFloatingEnvironment[
    fileext=loa,
    listname=List of Algorithms,
    name=Algorithm,
    placement=tbhp,
]{algorithm}
\usepackage[capitalize]{cleveref}
\crefname{figure}{Fig.}{Figs.}
\Crefname{figure}{Figure}{Figures}
\crefname{table}{Tab.}{Tabs.}
\Crefname{table}{Table}{Tables}
\crefname{equation}{Eq.}{Eqs.}
\Crefname{equation}{Equation}{Equations}
\crefname{section}{Sec.}{Secs.}
\Crefname{section}{Section}{Sections}
\crefname{paragraph}{Sec.}{Secs.}
\Crefname{paragraph}{Section}{Sections}
\crefname{algorithm}{Alg.}{Algs.}
\Crefname{algorithm}{Algorithm}{Algorithms}
\crefname{theorem}{Thm.}{Thms.}
\Crefname{theorem}{Theorem}{Theorems}
\crefname{corollary}{Cor.}{Cors.}
\Crefname{corollary}{Corollary}{Corollaries}
\usepackage{multirow}

\draft % marks overfull lines with a black rule on the right

\begin{document}

%\preprint{}



\title{A Parallel, Distributed Memory Implementation of the Adaptive Sampling Configuration Interaction Method}

\author{David B. Williams-Young}
\affiliation{Applied Mathematics and Computational Research Division, Lawrence Berkeley National Laboratory, Berkeley, California , USA}
\email{dbwy@lbl.gov}

\author{Norm M. Tubman}
\affiliation{NASA Ames Research Center, Moffett Field, California 94035, USA}

\author{Carlos Mejuto-Zaera}
\affiliation{Scuola Internazionale Superiore di Studi Avanzati (SISSA), Trieste TS, Italy}

\author{Wibe A. de Jong}
\affiliation{Applied Mathematics and Computational Research Division, Lawrence Berkeley National Laboratory, Berkeley, California , USA}


\date{\today}

\begin{abstract}
Many-body simulations of quantum systems is an active field of research that involves many different methods targeting  various computing platforms.  Many methods commonly employed, particularly coupled cluster methods, have been adapted to leverage the latest
advances in modern high-performance computing.
%take advantage of parallel computing and in some cases %methods can take advantage of such resources at a massive %scale.   
Selected configuration interaction (sCI) methods have seen extensive usage and development in recent years. However development of sCI methods targeting massively parallel resources has been explored only in a few research works.   In this work, we present a parallel, distributed memory
implementation of the adaptive sampling configuration interaction approach (ASCI) for sCI.  
In particular, we will address key concerns pertaining to the parallelization of 
the determinant search and selection, Hamiltonian formation, and the variational eigenvalue calculation 
for the ASCI method. Load balancing in the search step is achieved through the application
of memory-efficient determinant constraints originally developed for the ASCI-PT2 method. 
Presented benchmarks demonstrate parallel efficiency exceeding 95\%
for the variational ASCI calculation of Cr$_2$ (24e,30o) with $10^6, 10^7$, and $3*10^8$ variational
determinants up to 16,384 CPUs. To the best of the
authors' knowledge, this is the largest variational ASCI calculation to date.
%While there are several popular sCI variants, the ASCI approach is unique in that it 
%uses a set of constraint (triplet-constraints) to divide up the work for the perturbation theory steps.  
%These constraints have low memory requirements and naturally provides a way to divide up the work into many tasks.    In this work we not only extend the constraint approach to have even lower memory requirements, but in the process we achieve better load balancing for parallel task distribution while unifying the method for the search and perturbation theory steps.  These advances in conjunction with our approach to parallel diagonalization realizes a parallel selected configuration interaction approach specifically for the ASCI algorithm which we demonstrate in detail in this work.
\end{abstract}


\maketitle %\maketitle must follow title, authors, abstract and \pacs

\section{Introduction}
\label{sec:intro}

%Parallel many-body problem \cite{valeev21_many}
%Parallel CC \cite{solo2014}
The ability of \emph{ab initio} quantum chemistry to solve
challenging science problems strongly relies on its success in leveraging 
the latest advances in modern high-performance computing.
As such, the development of massively parallel algorithms 
for quantum chemistry methods targeting large supercomputers
has been of significant interest in recent years~\cite{bert1,janssen2008parallel,deJong2010utilizing,doi:10.1021/acs.chemrev.0c00700,Kothe19exascale,doi:10.1098/rsta.2019.0056}. In particular,
many-body methods hold among the highest potential for exploiting
concurrency on modern machines (we refer the reader to Ref.
\citenum{valeev21_many} and references therein for a 
recent comprehensive review). There are several competitive methods in quantum chemistry used for many-body simulation, including density matrix renormalization group (DMRG) \cite{schollwock2005,amaya2015}, coupled cluster\cite{bartlett2007,solo2014}, auxillary field quantum Monte Carlo \cite{booth2009,chang2016,Motta_WIREs_2018} and several others~\cite{szabo:book,Eriksen2020}.  It is an active field of research to understand the pros and cons of these various methods~\cite{Eriksen2020}, and almost all of these methods are being actively developed.

Recently, the development of selected configuration interaction
(sCI) methods has seen a renewed interest,
and there has been a push to adapt the approach to modern computing architectures. This interest
has been, in large part, driven by recent
advances in the core components of sCI
algorithms~\cite{evangelista2014,Tubman2016,holmes2016}, which have in turn extensively
broadened the reach of their applications
~\cite{garniron2017,Zimmerman2017,Coe2018,asci18_pt2,Lesko2019,Park2020,Mejuto2019,Mejuto2020,asci20_modern,asci20_casscf,Levine2020,Eriksen2020,Krem2020,Krem2021,Pineda2021,Goings2021,Bhatty2021,qchem1,Park2021,Dang2023,Li2018,Mejtuo2022,Herzog2022,JET2022,Chamaki2022,Yoffe2023,Coe2023,Wang2023}.
%Selected configuration interaction (sCI) algorithms have seen a lot of interest in recent years which is a result of several different nice properties of the algorithms, including ease of implementation and results that show high accuracy for many challenging problems in chemistry
%\DBWY{Maybe a block cite here?}.     
%Part of the new interest in sCI is due to recent advances in the core components of the algorithms~\cite{Tubman2016,holmes2016,evangelista2014}.  These advances started a push that extensively broadened the reach of sCI algorithms~\cite{garniron2017,Zimmerman2017,Coe2018,asci18_pt2,Lesko2019,Park2020,Mejuto2019,Mejuto2020,asci20_modern,asci20_casscf,Levine2020,Eriksen2020,Krem2020,Krem2021,Pineda2021,Goings2021,Bhatty2021,qchem1,Park2021,Dang2023,Li2018,Mejtuo2022,Herzog2022,Chamaki2022,Yoffe2023,Coe2023,Wang2023}.  
In particular, some of the newer applications include simulation of vibrational/nuclear degrees of freedom~\cite{Lesko2019,Bhatty2021}, dynamical mean field theory~\cite{Mejuto2019,Mejuto2020}, time evolution~\cite{Krem2020,Krem2021,Klymko2022}, Hamiltonian compression ~\cite{Chamaki2022}, new algorithms for quantum computing and benchmarking~\cite{Tubman2018,Yoffe2023,Kanno2023}, machine learning approaches~\cite{Pineda2021,Coe2018,Herzog2022}, and large-scale simulation of quantum circuits~\cite{Mull2023,Hirs2023}.   
While there has been a lot of interest in sCI, there are still a number of technical aspects that are open research problems within the realm of high-performance computing.  
The lack of parallel sCI implementations has been somewhat visible as in a recent comparison of different algorithms for the application of the benzene molecule~\cite{Eriksen2020}.   Several sCI methods were considered but none of them used anywhere close to the number of CPU hours that the most accurate methods required.  Although there has been at least one simulation of sCI using more than a  billion determinants~\cite{Li2018}, simulations of this scale have not been widely performed. This leads to the obvious question of what the limits of sCI approaches are when used in conjunction with modern high performance computing resources. 
In this work, we present a parallel,
distributed memory implementation of the
adaptive sampling configuration interaction
(ASCI) method \cite{Tubman2016} for sCI simulations.
%Given the broad range of applications described above, there is clear interest to adapt sCI methods to high performance computing.  

The ASCI algorithm was developed with the idea of making approximations to various aspects of sCI algorithms. The  original ASCI paper\cite{Tubman2016} included approximations to the search algorithm that reduced the number of search terms while maintaining highly accurate results.   Soon after the heatbath CI algorithm made further approximations to the search algorithm while still maintaining reasonable results in many systems~\cite{holmes2016}.  Initial results on Cr$_2$ suggested these sCI algorithms are competitive with the most accurate contemporary many body methods, including DMRG.  Many further developments increased the efficiency of different parts of the algorithm, including the development of a fast perturbation step with a deterministic constraint approach (ASCI-PT2)~\cite{asci18_pt2}.  In this current work, we look to turn the fast ASCI-PT2 algorithm into a massively parallel search step, while also developing a scaleable  parallel diagonalization step.  These make up the variational part of the sCI algorithm, and thus we are hoping to advance the parallel capabilities of variational sCI.

%\DBWY{I think we might want a high-level ASCI paragraph here - what is it, why is it adventageous, recent apps/sucesses, etc}

While single core and single node approaches for sCI have been explored extensively with a focus on making use of modern computational tools~\cite{asci20_modern}, there has been much less focus on massive parallelism and the development of distributed memory algorithms.  Most sCI implementations can make use of shared memory parallelism~\cite{asci20_modern,qchem1,Li2018} during the computationally heavy parts of the simulation.
In particular, leverage of multi-threaded~\cite{asci20_modern,solom,ips4}
and GPU accelerated~\cite{asci20_modern,asci18_pt2,thrust}
sorting techniques, which are the primary
algorithmic motifs of the ASCI method, have 
shown great promise on modern computing
architectures.
One recent study examined a
distributed memory implementation of the  
Heat-Bath configuration interaction
(HCI) method, particularly focusing on
the details pertaining to the strong
scaling of the second-order perturbative
corrections (PT2), which dominated their 
simulation. While this study was able to exhibit
excellent strong scaling for the PT2 step in
HCI, the challenges associated with this 
development are somewhat unique to HCI and
are not easily extended to other sCI
algorithms. THe ASCI-PT2 algorithm, which was first posted in 2018~\cite{asci18_pt2}, uses triplet constraints to facilitate the division of parallel tasks~\cite{asci18_pt2}.   Some of the new ideas in the recent parallel HCI~\cite{Dang2023} approach are used to overcome issues that are bottlenecks for only HCI PT2 approaches~\cite{Li2018}, and are not directly applicable to ASCI-PT2 with triplet constraints.  An even bigger issue that we address in this current work are the parallel bottlenecks in the variational part of an sCI algorithm, which appears to be a major bottleneck for all the sCI approaches we are aware of.   This is evident in the parallel HCI paper as their method exhibited
much less favorable strong scaling for
the variational component of the sCI calculation.
Thus the primary focus of the following developments
will be on the variational component of the ASCI algorithm,
of which a fair amount can be applied to other
sCI algorithms as well. %\DBWY{I'm thinking in

The remainder of this work will be organized as follows. In
\Cref{sec:asci} we review the salient features of the
ASCI method relevant to the development of parallel
algorithms. \Cref{sec:parallel_search} presents
a work partitioning and load balancing scheme relevant
to the development of scalable determinant selection
on massively parallel architectures. \Cref{sec:parallel_eigensolver}  addresses key considerations for the development of parallel algorithms for the solution of large-scale eigenvalue problems 
and the parallel construction of sparse Hamiltonian
matrices within the sCI formalism. We examine
the performance and scalability of the proposed
methods for a challenging test case (Cr$_2$) in
\cref{sec:results} and discuss future research
directions in \cref{sec:conclusions}.

\section{Methods}
\label{sec:methods}
The notation and common abbreviations used in this work are summarized in Table~\ref{tab:symbols}.

\begin{table}[]
\centering
\begin{tabular}{lp{60mm}}
\hline
\multicolumn{1}{|l|}{Symbol} & \multicolumn{1}{l|}{Explanation} \\ \hline
$\psi_k$ & The wave function in the current ASCI step\\
$C_i$ & The coefficients of the $i$th deteminant $D_i$ of $\psi_k$ \\
$D_{i}$ & The $i$th determinant in $\psi_k$ \\
$\tilde D_j$ & The $j$th determinant not in $\psi_k$ \\
%\{$C$\} & The set of coefficients in $\psi_k$ \\
%$\{D\}$ & The set of determinants in $\psi_k$ \\
\{$D^{sd}$\}& The set of all single and double excitations that are connected to $\psi_k$ \\
\{$D_{i}^{sd}$\} & The set of all single and double excitations that are connected to the determinant $D_{i}$ \\
%\{$D_{connect}^{sd}$\} & The set of all number pairs ($i$,$j$) such that $i$ is a determinant in $\{D\}$ and for a given $i$, the $j$ index must be in~\{$D_{i}^{sd}$\}. \\
$N_{tdets}$ & Number of determinants in the current wave function. \\
$N_{cdets}$ & Core space size used for pruning in ASCI\\
PE & Processing element (independent compute context)
\end{tabular}
\caption{List of symbols used in this work.}
\label{tab:symbols}
\end{table}

\subsection{Adaptive Sampling Configuration Interaction}
\label{sec:asci}
 
The details and efficacy of the ASCI algorithm are presented elsewhere. The initial implementation can be found in Ref.~\citenum{Tubman2016}, and many details and improvements can be found in Ref.~\citenum{asci20_modern}.
%in which we consider modern approaches that take advantage of different feature sets of CPUs and GPUs. 
The general schematic for the
ASCI algorithm is given in \cref{alg: ASCImain}. In general, the ASCI algorithm,
as most other sCI algorithms, consists of three major steps:
\begin{enumerate}
    \item Hamiltonian construction and diagonalization (eigenvalue problem),
    \item Determinant selection (search),
    \item (optional) Perturbative (PT2) correction to the energy.
\end{enumerate}
The first two steps are typically referred to as the variational steps for sCI
algorithm and will be the primary focus of this work.
However, in the development of parallel algorithms for the ASCI search step (\cref{sec:parallel_search}), we will extend a work partitioning scheme originally developed for the ASCI-PT2 method
\cite{asci18_pt2}. For the variational steps, sCI algorithms primarily differ in
their treatment of the determinant search, whereas once a set of determinants has been
selected, the Hamiltonian construction and subsequent diagonalization steps are
largely the same. As such, the parallel algorithms developed in this work for the latter
can straightforwardly be applied to other sCI algorithms as well. However, the details of 
the search step, which must be carefully considered in the development of scalable
parallel algorithms, are typically unique to each sCI algorithm, and thus not 
straightforwardly extended between methods.

\begin{algorithm}[t]
\begin{algorithmic}[1]
    \State \textbf{Input:} Start with a Hartree-Fock simulation 
    \State \textbf{Output:} Ground state energy of the system
    \State Create a starting wave function which can be a Hartree-Fock determinant
    \State ASCI-search (find important contributions not current in the wave function)
    \State Sort and select the top determinants from ASCI-search
    \State Diagonalize the Hamiltonian in the selected determinant space
    \State Return to step 3, but now use the ground state result of the diagonalization process as the starting wave function.  Repeat until stopping criteria is reached
    \State Run a final ASCI-PT2 step (to improve the energy further)
\end{algorithmic}
\caption{ASCI algorithm}
\label{alg: ASCImain}
\end{algorithm}
 
% The ASCI algorithm in particular was designed with unique algorithms for the search and PT2 subroutines that were optimized for modern CPUs/GPUs.   One of the key differences between different selected CI approaches is the way in which they approximate the search and PT2 steps.   
With the possible exception of a few approaches such as the machine learning sCI methods~\cite{Pineda2021,Herzog2022,Coe2018,Goings2021}, the vast majority of sCI algorithms use the following formula for generating a ranking value of a determinant not currently considered in the trial wave function, $\psi_k$,
\begin{equation}
    S_i = \sum_j S_i^{(j)}, \qquad S_i^{(j)} = \frac{H_{ij}C_j}{H_{ii} - E_k}, \qquad E_k = \langle\psi_k\vert H \psi_k \rangle.\label{eq:asci_score}
\end{equation}
%A similar expression for the PT2 energy correction $E_{PT2}$ was derived from Epstein and Nesbet ~\cite{epstein1926,nesbet1955} and is given as follows,
% \begin{equation}
% E_{PT2} =
% \sum_{i}\frac{
% \sum_{j} H{ij}C_{j}}{E_{ASCI(var)}-H_{ii}}. \label{eqn:en}
 %\sum_{i}\frac{|\langle\psi|H|D_{i}\rangle|^{2}}{E_{ASCI(var)}-H_{ii}}. \label{eqn:en}
% \end{equation}
%If we consider the contribution of each perturbation term, we see a very similar formula to the the search ranking,
 %\begin{equation}
 %E_{i} =
 %\frac{
 %\sum_{j} H{ij}C_{j}}{E_{ASCI(var)}-%H_{ii}}. \label{eqn:en-i}
 %\end{equation}
 Here, $H$ is the many-body molecular Hamiltonian, $H_{ij}$ is a matrix element of $H$ between Slater determinants, $C_j$ is the coefficient for $D_j\in\psi_k$,
 and $E_k$ is the variational energy associated with $\psi_k$. $S_i$ is the
 metric (``score") by which we will determine which determinants to add to the sCI subspace,
 and $S^{(j)}_i$ is the $j$-th partial-score which describes the contribution of $D_j$ to $S_i$. The contracted index, $j$, runs over
 $D_j\in\psi_k$ whereas the free index, $i$, runs over $\tilde D_i\not\in\psi_k$.
% The terms in these equations involve the Hamiltonian matrix elements between $i$ and $j$ determinants, H$_{ij}$, an input trial wave function with coefficients, $C$, a trial energy or a variational energy associated with the input wave functionm $E$.
In principle, $i$ runs over the entire Hilbert space in the complement of the determinants in $\psi_k$. However,
the elements of $H_{ij}$ are sparse, and the maximally two-body nature of $H$ for molecular
Hamiltonians imposes $\tilde D_i\in \{D^{sd}\}$, where $\{D^{sd}\}$ is the set of determinants
which are singly or doubly connected to a determinant in $\psi_k$. In practice, even consideration
of all determinants in $\{D^{sd}\}$ is not needed, as there exists considerable numerical sparsity beyond
that which is solely imposed by connections in the Hamiltonian.
%
%For both of these equations there is a sum over all current configurations in the current wavefunction $j$, and configurations $i$ that are not currently included in the wave function.  The number of terms in $j$ would be the entire Hilbert space, however if not for the sparse property of $H_{ij}$.  The terms in $H_{ij}$  are nearly zero for everything except of terms which are singly/doubly excited from the current wave function.  Since multiple terms in $i$ can be singly/doubly excited from a particular $j$, one has to find all these connections to properly calculate the ranking term.  
Thus, while calculating these equations is straightforward in principle, in practice it is important to identify the non-zero terms \emph{a priori} because the large majority of terms are zero.  Additionally, it is important to only use and store the non-zero terms when needed, as memory requirements for storing all the terms in these sums are extremely large when pushing to a large sCI calculation.   Previous papers have reported examples pertaining to the number of connections of different examples. In a previous ASCI study, the number of such terms considered for a Cr$_2$ example had over 282 billion  connection terms which were all generated on a single core~\cite{asci18_pt2} and in the recent parallel HCI paper, the authors were able to simulate over 89.5 billion connection terms.  If all of these terms had to be stored in memory simultaneously, the memory requirements would make these approaches unfeasible.

Due to the immense cost of calculating $H_{ij}$
and $S_i$, small differences in data structures
and algorithmic design choices can yield large
impacts on overall performance.
Since its inception, the design approach of the ASCI algorithm has centered around
 the use of modern computing architectures. In
 particular, the ASCI method has adopted sorting as its primary
 algorithmic motif due to the favorable scaling and performance of sorting 
 algorithms on contemporary CPU and GPU hardware\cite{asci20_modern}. However, current developments of the ASCI algorithm have focused on shared memory compute environments.   Almost all ASCI simulations that we are aware of have been performed on a single node with the exception of a small scale MPI test with parallel ASCI-PT2~\cite{asci18_pt2}.
 In the following subsections, we examine the extension of the ASCI sorting motif to distributed memory compute environments and emergent challenges with load balancing on massively parallel architectures.

\subsection{Parallel Determinant Search}
\label{sec:parallel_search}

%\DBWY{This needs to get worked in a bit better, might make sense ti merge with the subsequent paragraph} At high level, ASCI was designed to make use of well developed data structures and libraries that have favorable scaling properties on modern computers.   
%We can imagine the number of bit operations in a sCI algorithm to be \DBWY{you seem to like this number - but I think we would need justification to talk about it explicitly. Do you have a reference for something like this?} $10^{12}$ or more and thus not only are efficient algorithms are needed for the bit manipulation, but efficient movement of data is important.   To facilitate this, we tested extensively different  tree, hash tables, and sorting approaches in previous works.  The development of these libraries~\cite{boost2011,plauger2000} inherently require nearly optimal uses of memory and object comparisons, thus allowing us to focus on building ASCI on top of highly optimized algorithms.  After extensive benchmarking~\cite{asci20_modern} it was determined that  sorting is the fastest approach for bitstring management and manipulation across the entirety of the ASCI algorithm.   We note however that hash tables and tree approaches could easily be substituted in if such algorithms eventually become more efficient~\cite{hash_benchmark}.  Our tests also include tests on GPUs, which also showed sorting to be the most efficient of approaches. 

\begin{algorithm}[t]
\begin{algorithmic}[1]
\State \textbf{Input:} Trial wave function determinants, $\psi_k$ and coefficients $C$; Search configuration cutoff $N_{cdets}$; Max wave function size $N_{tdets}$; ASCI pair threshold, $\epsilon_{search}$
\State \textbf{Output:} New ASCI determinants $\psi_{k+1}$
\State $\psi_c \leftarrow $ Obtain
  $N_{cdets}$ subset of $\psi_k$ with
  largest coefficients
\State $P\leftarrow [\,\,]$\Comment{ASCI Pairs}
\For{$D_j\in\psi_c$}
  \State $\{D^{sd}_j\} \leftarrow$ Single and double connections from $D_j$
  \For{$\tilde D_i \in \{D^{sd}_j\}$}
     \State $S^{(j)}_i\leftarrow$ \cref{eq:asci_score}
     \If{$\vert S^{(j)}_i\vert > \epsilon_{thresh}$}
       \State $P\leftarrow [P,\,\, (\tilde D_i, S^{(j)}_i)]$
     \EndIf
  \EndFor
\EndFor
\State $P\leftarrow$ Sort $P$ on $\tilde D_i$.
\State $P\leftarrow \sum_{j} S^{(j)}_i$ for each unique $i$ \Comment{Each unique $p\in P$ now contains a complete $S_i$}
\State $P\leftarrow$ Sort $P$ on $S_{i}$
\State \Return $\psi_{k+1}\leftarrow$ top-$N_{tdets}$ determinants in $P$
\end{algorithmic}
\caption{The ASCI Search Algorithm}
\label{alg:asci_search}
\end{algorithm}

The general procedure for the ASCI
determinant search is given in
\cref{alg:asci_search}. The search proceeds by considering single and double connections from a set of dominant configurations in the trial wave function \cite{asci20_modern}. We will refer to these determinants as the ``core" determinants in the following.
The number of core determinants to consider is a tunable parameter in the ASCI algorithm and is typically achieved by
selecting the $N_{cdets}$ determinants in $\psi_k$ with the largest 
coefficient magnitudes. In a naive approach, one determines \emph{a priori}
the $\{D^{sd}\}$ associated with the core determinants, which
we will refer to as the \emph{target} determinants in the following, 
and subsequently calculates the contribution of each core determinant 
to each target configuration via \cref{eq:asci_score}. In many practical simulations, it is better to generate the possible target determinants as singles and doubles from each core configuration, rather than loop over all possible target configurations and then check for connections with the core determinants.  The latter 
procedure generally performs excessive work as the majority of the 
core determinants only contribute to a few target determinants.
In the shared memory ASCI algorithm, a single pass is taken over the
core determinants, and the associated $\{D^{sd}_j\}$ and $S_i^{(j)}$ for
$\tilde D_i\in\{D^{sd}_j\}$ are generated on the fly. If $\vert S_i^{(j)}\vert > \epsilon_{search}$, where $\epsilon_{search}$ is
a tunable threshold,
it is appended to an array as a pair $(\tilde D_i,S_i^{(j)})$,
which we will refer to as an ASCI pair in the following. For the purposes of molecular
calculations, bit-strings are typically the data structure of choice for the representation and manipulation
of many-body determinants as it allows for fast bit operations in e.g. calculating matrix elements on modern computing architectures
\cite{spencer2019,qchem1,asci20_modern,evangelista2014,Dang2023}.   As such, the ASCI pairs can be stored in a simple
data structure consisting of the bitsring representing $\tilde D_i$ and the partial score $S_i^{(j)}$.
Once all pairs
have been generated after passing over each core configuration, the
array is sorted on the determinant bitstring which in turn brings all partial
scores associated with a particular target determinant contiguously in memory. 
With this new ordering, scores for individual target determinants can be straightforwardly
assembled via \cref{eq:asci_score}. We refer to this procedure as ``sort-and-accumulate"
in the following.

Even for relatively aggressive
$\epsilon_{search}$ cutoffs, this approach
can constitute a large memory footprint for
larger $N_{cdets}$. In the original ASCI method,
a threshold for the maximum number
of ASCI pairs to store in memory was set, and
the pair list would be occasionally pruned to ensure memory compactness. 
This pruning procedure has been demonstrated to provide approximate scores
of sufficient accuracy to successfully rank determinants for the sCI wave function\cite{asci20_modern}.
A similar approach could be taken for the calculation of the PT2 energy, $E_{PT2}$,
which is closely related to the scores of \cref{eq:asci_score}~\cite{asci18_pt2,Li2018,garniron2017}.
However, for accurate calculations of $E_{PT2}$, one cannot employ the aggressive
cutoffs and pruning schemes employed to ensure memory compactness in the ASCI
search. In the development of the ASCI-PT2 method\cite{asci18_pt2}, a solution was developed
which allows for memory compactness while still leveraging the powerful
nature of sorting algorithms on modern computing architectures. In this procedure,
the set of target configurations is partitioned according to a constraint
identified by their largest occupied orbitals. In the ASCI-PT2 study, target
determinants were classified by triplet constraints, i.e., by their three highest-occupied
orbitals. 
These constraints generate non-overlapping subsets of determinants such that
every possible determinant belongs to exactly one constraint class.

At first glance, this approach appears similar to the naive approach
for ASCI score calculation as it might imply that one would need to know the entire
$\{D^{sd}\}$ associated with the core determinants to be able to partition
the target determinants \emph{a priori}. However, the major realization of ASCI-PT2
was to determining which target determinants arising from excitations of a particular
core configuration belong to a particular constraint class. As such, in a similar
manner to the original ASCI search algorithm, one can make multiple passes over the core
configurations for each constraint and generate its associated target configurations 
on the fly. 
%The pseudocode for generating all the connecting determinants belonging to a particular constraint is given in Algorithm~\ref{alg:constsingles} for the single connections and Algorithm~\ref{alg:constdoubles} for the double connections. In these algorithms, the $\vert \cdot \vert$ denotes a count operation which returns how many mask values are set (i.e., a \texttt{popcount}).
The power of this procedure is that the sort-and-accumulate steps can now
be done at the constraint level, i.e., ASCI pairs can be generated and aggregated for
each constraint and, due to the non-overlapping partition generated by the individual
constraints, all partial scores associated with the target determinant belonging
to that class are guaranteed to be generated by the end of each pass over core
configurations. As the number of unique determinant scores is much less than
partial score contributions, this procedure exhibits much lower memory 
usage than the original ASCI search algorithm.

\begin{algorithm}[t]
\begin{algorithmic}[1]
\State \textbf{Input (global): } Dominant core configurations, $\psi_c$ , max constraint level $L$ 
\State \textbf{Output (local):} Local constraints $C_{loc}$
\State $W \leftarrow$ an array of size of the execution context (\# PEs) \Comment{Initialize to zero}
\State $p\leftarrow$ index of PE in 
\State $C_{loc}\leftarrow[\,\,]$
\State $C_t\leftarrow$ all unique triplet constraints
\For{$C\in C_t$}
\State $h\leftarrow 0$
\For{$D_j\in\psi_c$}
\State $s\leftarrow $ singly connected determinants to $D_j$ satisfying $C$. %\Comment{\cref{alg:constsingles}}
\State $d\leftarrow $ doubly connected determinants to $D_j$ satisfying $C$.
%\Comment{\cref{alg:constdoubles}}
\State $h\leftarrow h + \vert s \vert + \vert d\vert$
\EndFor
\If{$h>0$}
\State $q\leftarrow \arg\min_i W_i$
\If{$p=q$}
\State $C_{loc}\leftarrow [C_{loc},\,\, C]$
\EndIf
\EndIf
\EndFor
\State \Return $C_{loc}$
\end{algorithmic}
\caption{Load Balancing for Parallel Constraint ASCI Search}
\label{alg:search_loadbalance}
\end{algorithm}

%For both the PT2 and search algorithms we are looking to calculate a quantity for bit strings currently not included in our current wave function, but has a hamming distance less than 4 from at least one bitstring currently included in the current wave function.  For the search algorithm, we are looking to calculate an amplitude for each bitstring, for the perturbation theory algorithm, we are looking to calculate an energy contribution to each bit string.    The ASCI search and perturbation theory algorithms allow for approximate and exact calculcation of these quantities, within a structure that trivially parallelizes.  
To proceed with development of a parallel ASCI search algorithm we consider a similar approach to ASCI-PT2, where the work associated with the generation of pairs for
individual constraints is assigned to different processing elements (PE) in 
the distributed memory environment. To ensure scalability of the pair
generation and subsequent sort-and-accumulate steps, we have developed
a static load balancing procedure presented  in \cref{alg:search_loadbalance}. Prior to any pairs being generated for the ASCI search, 
the number of determinants associated with individual constraints
is precomputed on the fly to generate the rough amount of work for each constraint.
The number of contributions is then used as a heuristic to assign work to individual PEs. It's important to note that the number of contributions is not
an exact measure of the work associated with a particular constraint due
to the fact that contributions are screened out according to $\epsilon_{search}$. However, for the purposes of the ASCI search,
this procedure has been demonstrated to generate sufficiently balanced work (see \cref{sec:parallel_search}).
The primary challenge of this work distribution scheme
stems from the large variance of work associated with any particular constraint, i.e.
it is often the case that a few constraints yield a disproportionate
of target determinants. As such, at a particular level of constraint
(e.g. triplets, etc), there always exists an amount of indivisible work
that the load balancing procedure cannot distribute over PEs. The 
work associated with a particular constraint can be further divided
by considering addition constraints of higher order, e.g. triplets
can be broken up into quadruplets, quadruplets into quintuplets, etc.
However, the number of constraints at a particular level grows polynomially
with the number of single particle orbitals, e.g. $O(n_{orb}^3)$ for triplets,
 $O(n_{orb}^4)$ for quadruplets, etc. Due to the fact that for each constraint
 there is an inner loop over $N_{cdets}$ core configurations, it is
 highly desirable to limit the number of considered constraints.  
 As such,
 it would be highly inefficient to simply consider all higher-level constraints if the work associated with a particular constraint level
 is deemed to exhibit excessive load imbalance. Instead, it is advantageous
 to only divide the constraints which are expected to exhibit excessive
 work into higher constraint levels rather than dividing all constraints.
 We will demonstrate the efficacy of this partitioning scheme in \cref{sec:parallel_search}.

\begin{algorithm}[h]
\begin{algorithmic}[1]
\State \textbf{Input:} Local part $A_I$ of an array $A$ with $\vert A \vert = n$, Element index $k\in [1,n]$
\State \textbf{Output:} The element $a_k$ (global) such that $\vert \{ a > a_k\,\vert a\in A\} \vert = k$
\State \Comment{All operations are local unless otherwise specified.}
\State  $rank\leftarrow $ PE rank.
\State  $n_I \leftarrow \vert A_I \vert$  
\State  $N \leftarrow \mathrm{Allreduce}(n_I)$ \Comment{Collective}
\While{$N \geq n_{min}$}
  \State $p \leftarrow$ Select a random pivot in $[0,N)$.
  \State Determine the owner of $a_p$ and broadcast to other PEs \Comment{Collective}
  \State  $A^g_I, A^e_I, A^l_I \leftarrow \mathrm{partition}(A^I,a_p)$ 
  \State  $G_I, E_I, L_I \leftarrow \vert A^g_I\vert, \vert A^e_I\vert, \vert A^l_I\vert$ 
  \State  $\{G,E,L\} \leftarrow \mathrm{Allreduce}(\{G_I, E_I, L_I\})$ \Comment{Collective}
    \If{$k\leq G$}
      \State $A_I\leftarrow A^g_I$
    \ElsIf{$k\leq G+E$}
      \State $a_k\leftarrow a_p$ \\
      \Return $a_k$
    \Else
      \State $A_I\leftarrow A^l$ 
      \State $k\leftarrow k - G - E$ 
    \EndIf
    \State  $n_I \leftarrow \vert A_I \vert$ 
  \State  $N \leftarrow \mathrm{Allreduce}(n_I)$ \Comment{Collective}
\EndWhile
\State $A_{rem} \leftarrow \mathrm{Gather}(A_I)$ \Comment{Collective (PE 0)}
\If{$rank = 0$}
 \State  $a_k\leftarrow \mathrm{SerialQuickselect}(A_{rem})$ \Comment{\texttt{std::nth\_element}}
\EndIf
\State Broadcast $a_k$ \Comment{Collective}\\
\Return $a_k$
%  \caption{Distributed Memory Quickselect}
%  \label{alg:distquickselect}
  %\SetKwInOut{kinput}{Input}
  %\SetKwInOut{koutput}{Output}
  %\SetKwRepeat{Do}{do}{while}

  %\kinput{Local part $A_I$ of an array $A$ with $\vert A \vert = n$, Element index $k\in [1,n]$}
  %\koutput{The element $a_k$ (global) such that $\vert \{ a > a_k\,\vert a\in A\} \vert = k$} 
 % \begin{algorithmic}
 % \Loop{
 %   \State  $n_I \leftarrow \vert A_I \vert$ (local)
 %   \State  $N \leftarrow \mathrm{Allreduce}(n_I)$ %(collective)
%    \If{$N\leq n_\mathrm{max}$}
%      \State \textbf{break}
%    \EndIf
%    \State  $p \leftarrow$ Select a random pivot in $[0,N)$. (local)
%    \State  Determine the owner of $a_p$ and broadcast to other PEs (collective)
%    \State  $A^g_I, A^e_I, A^l_I \leftarrow \mathrm{partition}(A^I,a_p)$ (local)
%    \State  $G_I, E_I, L_I \leftarrow \vert A^g_I\vert, \vert A^e_I\vert, \vert A^l_I\vert$ (local) 
%    \State  $\{G,E,L\} \leftarrow \mathrm{Allreduce}(\{G_I, E_I, L_I\})$ (collective)
%    \If{$k\leq G$}
%      \State $A_I\leftarrow A^g_I$
%    \ElsIf{$k\leq G+E$}
%      \State $a_k\leftarrow a_p$ 
%      \State \textbf{break}
%    \Else
%      \State $A_I\leftarrow A^l$ 
%      \State $k\leftarrow k - G - E$ 
%    \EndIf
%  }
%  \end{algorithmic}
\end{algorithmic}
\caption{Distributed Memory Quickselect}
\label{alg:dist_quickselect}
\end{algorithm}

Given the locally constructed arrays of string-score pairs generated as a result
of the constraint partitioning, the remaining aspect of the parallel ASCI search
is to determine the dominant $N_{tdets}$ determinants to keep for the next iteration.
In principle, this can be achieved by performing a full sort of the pairs on the absolute
value of their scores. For an array $A$ such that $\vert A\vert = n$, full sorting of $A$ scales
$O(n\log{n})$, which is generally unfavorable for the large sizes of arrays considered in this
work (as would be generated by large values of $N_{cdets}$). For distributed memory
implementations, this problem would be exacerbated due to the communication overhead of existing distributed memory sort algorithms\cite{solom,10.1145/2755573.2755595}.
For the vast majority of properties involving selected-CI wave functions, such as energies and
density matrices, the \emph{order} in which the determinant coefficients appear in the CI vector is
 irrelevant and all computed quantities will be invariant to arbitrary permutations 
$C\leftarrow PC$. 
%(however, see \cref{sec:parallel_eigensolver} for examples how ordering might affect the performance of some parallel algorithms). 
As such, the sorting problem can 
be replaced with the \emph{selection} problem which can be used to determine the largest 
(or smallest) $k \leq n$ values of an array with $O(n)$ complexity ~\cite{selection1,selection2}. In addition,
selection algorithms can be performed in parallel with nearly optimal speedup without
having to communicate significant segments of $A$ \cite{selection3}.
Rather than obtaining an absolute ordering of its input data, 
selection algorithms are designed to determining a \emph{pivot}, $a_k \in A$, such that
$\vert A^g \vert = k$, where $A^g = \{a > a_k\,\vert\,a\in A\}$. In cases where $a_k$ appears
multiple times in $A$, this definition can be extended to indicate 
$\vert A^g \vert < k \leq \vert A^g \cup A^e \vert$, where $A^e\subset A$ is the indexed
subset containing the dubplicate elements of $a_k$.
 We outline a duplicate-aware, distributed memory extension of the \texttt{quickselect} 
 algorithm, with expected $O(n)$ runtime, in \cref{alg:dist_quickselect}. For the 
 ASCI search problem, \cref{alg:dist_quickselect} is used to determine the contribution
 pair with the $N_{tdets}$-largest score, $p_{tdets}$. We may then determine 
 the contribution pairs with scores larger-than or equal-to $p_{tdets}$ and subsequently 
 gather them (collectively) to all participating PEs (via e.g. \texttt{MPI\_Allgather(v)}).
 We examine the performance of this scheme, which introduces the only source of distributed
 memory communication in the presented ASCI search method, in \cref{sec:results}.

\subsection{Parallel Eigensolver}
\label{sec:parallel_eigensolver}
%NMT: I will avoid working on this section until we speak further.  Its the hardest section for me
%\begin{itemize}
%    \item Parallel SPMV
%    \item Parallel Hamiltonian Construction
%    \item Parallel Davidson/LOBPCG method
%\end{itemize}
%Roel's LOBPCG papers
%Chao's LOBPCG paper\cite{ming18_robust}
%CA Krylov\cite{ca_krylov}
%Parallel Krylov\cite{saad89_krylov}

%Parallel Davidson\cite{oliveria98_parallel,nool00_parallel,romain14_parallel,romero10_parallel,stathopoulos93_reducing}

After each search iteration of the ASCI procedure, we must
obtain the ground state eigenpair of the many-body Hamiltonian projected onto the
basis of newly selected determinants.
The large basis dimensions ($N_{tdets}$) employed in accurate sCI applications precludes
the use of direct eigensolvers (e.g. those implemented in dense linear algebra libraries
such as (Sca)LAPACK \cite{scalapak}~ and ELPA \cite{elpacode,elpa1}) due to their 
 $O(N^2)$ dense storage requirement
and steep $O(N^3)$ scaling with problem dimension. As such, Krylov-subspace methods, such as 
Davidson \cite{davidson1975}, LOBPCG \cite{lob1,lob2}, and Lanczos \cite{lan1}, are typically employed. 
Development of efficient and scalable Krylov methods on
distributed memory computers
is challenging due to the existence of many synchronization points arising from the serial nature of subspace construction. 
Significant research effort has been afforded to the development of distributed memory Krylov eigenvalue methods across many 
scientific computing domains \cite{saad89_krylov,stathopoulos93_reducing,oliveria98_parallel,nool00_parallel,romero10_parallel,romain14_parallel,ming18_robust,yang20_scalable,yang22_enhancing, ca_krylov}. In this
work, due to the diagonally dominant nature of the Hamiltonian,
we consider the parallel implementation of the 
Davidson method for sCI applications, although many of the
same principles can be extended to other Krylov methods such
as Lanczos or LOBPCG.

\begin{algorithm}
\begin{algorithmic}[1]
    \State \textbf{Input:}  Matrix-vector product operator, \texttt{OP}; preconditioner operator \texttt{K}, max Krylov dimension $k$; initial guess $v_0$; Residual norm tolerance $\epsilon$.
    \State \textbf{Output:} Eigenpair $(E_0,v)$ approximating the lowest eigenvalue of \texttt{OP}.
    \State \Comment{All vectors are distributed among PEs}
    \State $w \leftarrow \texttt{OP}(v_0)$
    %\State $v \leftarrow \texttt{PORTH}(v_0, w)$ \Comment{Parallel Gram-Schmidt}
    \State $h\leftarrow v_0^\dagger w$; $h\leftarrow \mathrm{Allreduce}(h)$; $v \leftarrow w - v_0h$ \Comment{Parallel Gram-Schmidt}
    \State $W\leftarrow w$; $V\leftarrow [v_0\,\,v]$
    \State $i \leftarrow 1$
    \While{$i<k$}
      \State $W \leftarrow [W\,\, \texttt{OP}(v)]$
      \State $H_{loc} \leftarrow W^\dagger V$ \Comment{Local all PEs}
      \State $H_{tot} \leftarrow \texttt{Reduce}(H_{loc})$ \Comment{Collective to PE 0}
      \State Solve $H_{tot}\tilde C = \tilde C \tilde E$ \Comment{Local on PE 0}
      \State Broadcast $(\tilde E,C)$ \Comment{Collective}
      \State $R \leftarrow (W - \tilde E_0 V)\tilde C_0$ \Comment{Local all PEs}
      \State $r_{loc}\leftarrow R^\dagger R$ \Comment{Local all PEs}
      \State $\lVert r\rVert\leftarrow \sqrt{\texttt{Allgather}(r_{loc})}$ \Comment{Collective}
      \If{$\lVert r\rVert \leq \epsilon$}
        \State \Return $(E_0, v) \leftarrow (\tilde E_0, V\tilde C_0)$
      \Else
        %\State $v\leftarrow \texttt{PORTH}(V, R)$ \Comment{Parallel Gram-Schmidt}
        \State $R\leftarrow\texttt{K}(R)$ \Comment{Preconditioned Residual}
        \State $h\leftarrow V^\dagger R$; $h\leftarrow \mathrm{Allreduce}(h)$; $v \leftarrow R - Vh$ \Comment{Parallel Gram-Schmidt}
        \State $V\leftarrow[V\,\, v]$
      \EndIf
      \State $i \leftarrow i + 1$
    \EndWhile
\end{algorithmic}
\caption{Parallel Davidson Algorithm}
\label{alg:par_davidson}
\end{algorithm}

\begin{figure}[tbh]
    \centering
    \importlocalfigure{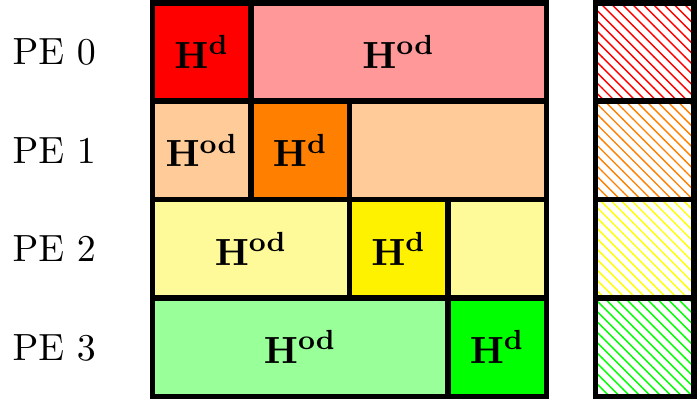}
    \caption{Example distributed memory storage scheme for the Hamiltonian and Davidson trial vectors across 4 PEs. The diagonal
    ($\mathbf{H^d}$, darker) and off-diagonal ($\mathbf{H^{od}}$, lighter) blocks of $H$ are
    stored separately on each PE as CSR matrices. For PEs 1 and 2, the off-diagonal blocks are stored as a single sparse matrix with elements on both sides of the diagonal block. Vectors (textured) are distributed according to the same row
    distribution as $H$. 
    }
    \label{fig:mat_vec_dist}
\end{figure}

Given an algorithm to produce the action of the Hamiltonian
onto a trial vector ($\sigma$ formation), we present 
the general 
scheme for the distributed memory Davidson method in
\cref{alg:par_davidson}. The algorithm presented is general to an arbitrary preconditoner, although we will always adopt the shifted Jacobi (diagonal) preconditioner which has become standard in configuration interaction calculations \cite{davidson1975}. 
Unlike exact diagonalization configuration
interaction methods (e.g. full/complete active space CI),
where implicit $\sigma$ formation exhibits efficient 
closed-form expressions~\cite{bert1,spencer2019}, the vast majority of 
sCI methods require explicit construction of $H_{ij}$ to perform
the necessary contractions. We refer the reader to 
Ref. \citenum{asci20_modern} for a comprehensive discussion
of efficient algorithms for the assembly of $H_{ij}$ 
for sCI wave functions. Although explicit storage of the Hamiltonian can be avoided via recalculation of the
$H_{ij}$ at each Davidson iteration, the immense cost
of matrix element construction renders this procedure 
prohibitive for practical applications. Sparsity of the the Hamiltonian in the basis of determinants allow for its explicit storage using 
sparse matrix storage formats. In
this work, we will use the distributed sparse matrix format
depicted in \cref{fig:mat_vec_dist}, which has been employed
for the development of parallel Krylov algorithms in many
community software packages for sparse matrix algebra (e.g. Ref.\citenum{petsc-user-ref}). 
In this storage
scheme, each PE owns a contiguous (not necessarily equal) 
row subset of the full
sparse matrix divided into diagonal and off-diagonal blocks
which are stored in standard sparse matrix formats. We will use the compressed sparse row (CSR) 
format for local matrix storage \cite{sparsematrix}. 
The partitioning into diagonal and off-diagonal blocks will ultimately determine the communication
pattern of the $\sigma$ formation in the Davidson method,
as will be apparent in \cref{sec:parallel_eigensolver}.
Of the
several methods for matrix element construction
discussed in Ref. \citenum{asci20_modern}, we consider the double loop approach in this work.  In the limit that each task only has a small portion of the Hamiltonian, the double loop approach is quick at generating all the matrix elements.   There is likely time savings to be had by optimizing this further with consideration to other approaches of generating the matrix elements like with residue arrays and dynamic bitmasking~\cite{asci20_modern}.  However, the cost analysis must take into account inefficiencies due to load balancing.  Thus a double loop approach is quite reasonable in terms of load balancing and we consider it a reasonable starting point for massively parallel approaches to sCI.
%\DBWY{NORM},  \DBWY{Justify double loop for dist H construction }

\begin{algorithm}
\begin{algorithmic}[1]
    \State \textbf{Input:} Matrix $H$ and vector $v$ in block distributed format (\cref{fig:mat_vec_dist}). (Optional) Precomputed communication data (\cref{eq:spmv_comm})
    \State \textbf{Output:} $w = Hv$ in conforming row partitioned format
    \State
    \If{Communication data needed}
       \State Compute communication data for $H$ per \cref{eq:spmv_comm}
    \EndIf
    \State Post local receives for incoming data from remote PEs \Comment{\texttt{MPI\_Irecv}}
    \State Pack elements of $v$ required by remote processes \Comment{Local}
    \State Post local sends to satisfy remove receive requests \Comment{\texttt{MPI\_Isend}}
    \State $w\leftarrow H^{d}v$ \Comment{Local}
    \State Wait for remote receives to complete
    \State $v_{rem}\leftarrow$ Unpack remote data
    \State $w\leftarrow w + H^{od}v_{rem}$
\end{algorithmic}
\caption{MPI-Based Non-Blocking Parallel Sparse Matrix-Vector Product (SpMV)}
\label{alg:dist_spmv}
\end{algorithm}

A naive approach for the distributed $\sigma$ formation
would replicate the basis vectors on each PE such that
local sparse matrix-vector (SpMV) product could be done
without communication. However, in the context of a Krylov
method such as Davidson, this would require gathering the
locally computed elements of the SpMV to \emph{every} PE
at each iteration. For large problem sizes, the connectivity 
of the full gather operation would become a bottleneck.
Due to the sparsity of $H$, the computation of individual
elements of $\sigma$ does not require knowledge of the
entire basis vector; in fact, the number of elements 
required for the contraction of \emph{any} row of $H$
is (pessimistically) bounded by the number of single and double exciations
allowable from the single particle basis. As such, replication
(or communication) of the entire vector is not generally required. To minimize data communication
in the Davidson iterations,
we employ the vector partitioning scheme also depicted in
\cref{fig:mat_vec_dist}. Using the sparsity pattern
of the off-diagonal blocks of $H$, we can efficiently 
determine which elements of remote vector elements
need to be communicated between PEs as the union of the
row sparsities, i.e.
\begin{equation}
    \mathcal{C}_I = \bigcup_{i\textrm{ on PE }I}  \mathcal{R}_i,\qquad
    \mathcal{R}_i = \{ j\,\vert H^\mathrm{od}_{ij} \neq 0\}. \label{eq:spmv_comm}
\end{equation}
From this set, the owners of remote matrix elements can be
looked-up using the row distribution of $H$. Another benefit of this precomputation is that it allows 
for the overlap of 
communication and computation \cite{10.1093/oso/9780198788348.003.0004}. Via this 
distribution, the diagonal SpMV can be done without communication as the vector elements for the local SpMV reside
on the same PE by construction. Using non-blocking communication
primitives in the MPI library, communication can be initiated prior to starting the diagonal SpMV and finalized only once the remote elements are required for the off-diagonal SpMV. The pseudocode
for this distributed SpMV employed in this work is given
in \cref{alg:dist_spmv}.

\section{Results and Discussion}
\label{sec:results}

In this section we examine the strong scaling behaviour of the proposed
algorithms for Cr$_2$ / def-SVP\cite{weigend2005a} in an active space of (24e, 30e). This
test system is one of the challenge systems that has effectively been solved to chemical accuracy, but only in  the last 10 years~\cite{amaya2015,Tubman2016}.  Therefore Cr$_2$ is a  a good benchmark systems for testing and 
 developing methods modern method.   Molecular integrals were
obtained using the Molpro \cite{doi:10.1002/wcms.82,doi:10.1063/5.0005081} program and all ASCI calculations were
performed on the Haswell partition of the Cray XC40 Cori supercomputer at the
National Energy Research Scientific Computing Center (NERSC). Each Cori-Haswell
node consists of 2x Intel Xeon Processor E5-2698 v3 16-core CPUs with 128GB
DDR4 RAM. All jobs were run with 32 PEs (MPI ranks) per node.
%The software implementation of the parallelASCI method used in this work is available as open-source software. 
Energies for Cr$_2$ calculations presented in this 
work are given in \cref{tab:cr2_energies}. All calculations were performed using $\epsilon_{search}=10^{-10}$.
Although not the primary focus of this work, we note that the variational energy determined 
at $N_{tdets}=3*10^8$ is only 0.3 mEh higher in energy than the 
most accurate variational DMRG result for the same system in Ref.~\citenum{olivares2015ab}. %As such, we can say
%with high confidence that the ASCI procedure is convergent
%out to this scale 
%\DBWY{Is there a better way to put this?}.

%Calculations for $N_{tdets}=3*10^8$ and $N_{cdets}=100k$ were not run as this test case was used only to demonstrate the strong scaling behaviour of the overall ASCI method. 
%As we will demonstrate in the following, the search step scales primarily with $N_{cdets}$, not with $N_{tdets}$, and therefore the data presented for the strong scaling of the ASCI search is representative of all tests considered. 
In the following, we present strong scaling results for the parallel
ASCI methods proposed in this work.

\begin{table}[tbh]
    \centering
    \begin{tabular}{l|cc}
\multicolumn{1}{c}{} & \multicolumn{2}{c}{$N_{cdets}$} \\
        \hline
        $E_0 / $ Eh          & 100k        & 300k        \\
        \hline\hline
        $N_{tdets}=10^6$      & -2086.40966 & -2086.41018 \\
        $N_{tdets}=10^7$      & -2086.41769 & -2086.41781 \\
        $N_{tdets}=3*10^8$    & --          & -2086.42042 \\
        \hline\hline
    Reference DMRG\cite{olivares2015ab} & \multicolumn{2}{c}{-2086.42077} \\
    \hline
    \end{tabular}
    \caption{ASCI Ground State energies for Cr$_2$ (24e,30o)}
    \label{tab:cr2_energies}
\end{table}

\subsection{ASCI Search Performance}
\label{sec:res_search}

\begin{figure}
  \centering
  \begin{minipage}{0.33\textwidth}
    \importlocalfigure{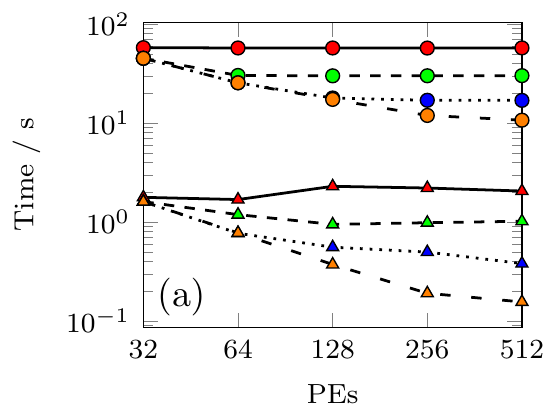}
  \end{minipage}~
  \begin{minipage}{0.33\textwidth}
    \importlocalfigure{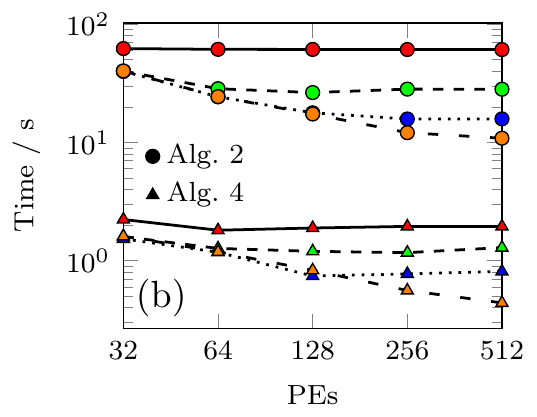}
  \end{minipage}~
  \begin{minipage}{0.33\textwidth}
    \importlocalfigure{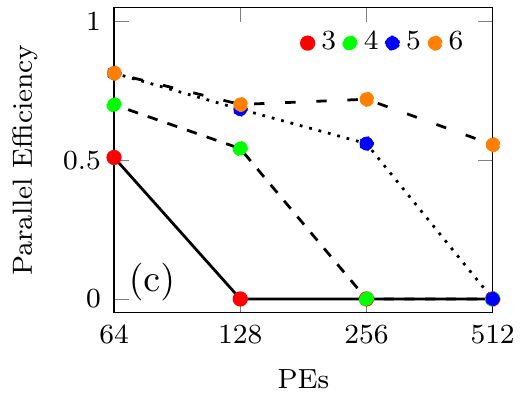}
  \end{minipage}
  \caption{Strong scaling of ASCI search components with
  different determinant constraints (triplet, 3; quadruplet, 4; qunituplet, 5; hextuplet, 6) for $N_{cdets}=300k$ and (a) $N_{tdets}=10^6$  (b) $N_{tdets}=10^7$. 
  Dominant components exhibit improved scaling with the addition of larger determinant constraints while
  lesser components are largely unaffected. (c) Illustrates the parallel efficiency of the overall ASCI search algorithm with different constraints.}
  \label{fig:search_scaling}
\end{figure}

\begin{figure}
  \centering
  \begin{minipage}{0.48\textwidth}
    \importlocalfigure{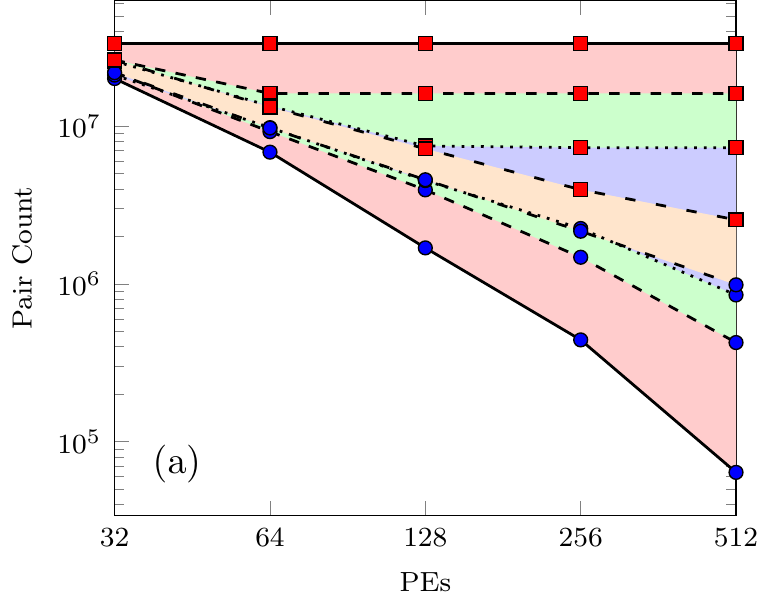}
  \end{minipage}~
  \begin{minipage}{0.48\textwidth}
    \importlocalfigure{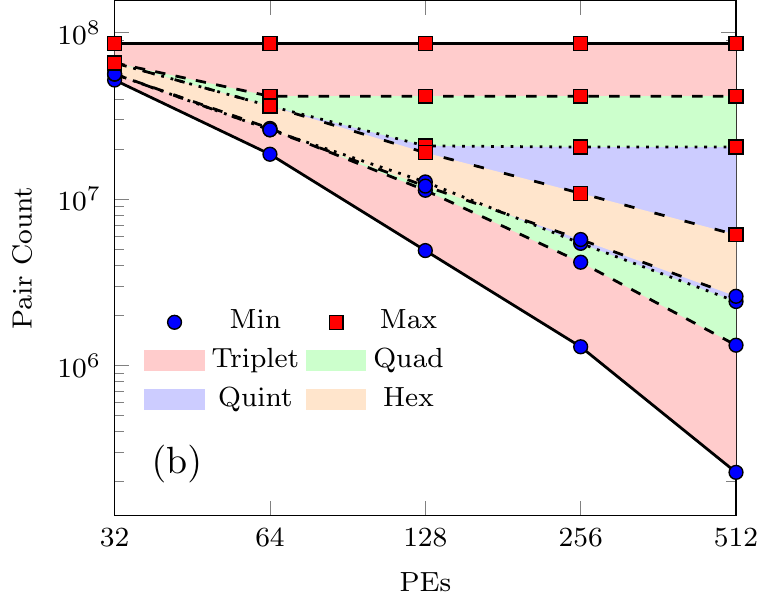}
  \end{minipage}
  \caption{Constraint pair count statistics for Cr$_2$ with $N_{tdets}=10^7$ and
  (a) $N_{cdets}=100k$ (b) $N_{cdets}=300k$. Triplet (3, red shading), quadruplet (4, green shading), quintuplet (5, 
  blue shading), and hextuplets (6, orange scaling)
  constraints are considered. A smaller shaded volume indicates better strong scaling of
  the constraint-partitioned ASCI search.}
  \label{fig:10Mcscaling}
\end{figure}

In \cref{fig:search_scaling}, we present the component and overall strong
scaling of the ASCI search for triplet, quadruplet, quintuplet, and hextuplet
determinant constraints (\cref{sec:parallel_search}) for Cr$_2$ with $N_{cdets}=300k$ and
$N_{tdets} = 10^6, 10^7$.  The calculation is dominanted by the generation of
ASCI pairs whereas the distributed determinant selection via \cref{alg:dist_quickselect} is relatively
minor in comparison ($<$ 1s). As such, the execution time of the ASCI search is largely independent of $N_{tdets}$. At the lowest determinant constraint level
(triplets), the presented algorithm stagnates immediately with no further
performance improvement with large PE counts. Strong scaling is improved
significantly by using larger determinant constraints. Overall,
we see a 5-5.5x performance improvement in the strong scaling limit by moving
from triplet to hextuple determinant constraints. The reason for this
improvement (or conversely why the triplet case exhibits no strong scaling at
all) can be seen in \cref{fig:10Mcscaling} where we present a plot showing the
effect of constraint size on the number of ASCI pairs generated per-PE.  At the
triplet level, there is a single triplet which yields a disproportinate number
of pair contributions. As the work can only be distributed at the level of the
highest determinant constraint, this single triplet is represents indivisible
work.  By breaking up the offending triplet into higher constraints, the work
is able to be distributed. However, at each constraint level, there always
reaches a point where there is a disproportionate amount of work for a few
particular constraints.  As such, we expect there always to be a strong scaling
stagnation for this procedure, but the inclusion of higher constraints
(heptuples, octuples, etc) will continue to improve the strong scaling
behaviour at larger PE counts. However, due to the relatively low cost of
the search relative to the Hamiltonian formation and Davidson iterations (see
\cref{sec:res_ham_eig}), we believe this level of optimization to be
unnecessary at this time.

\subsection{Hamiltonian and Eigensolver Performance}
\label{sec:res_ham_eig}

\begin{figure}
  \centering
  \begin{minipage}{0.33\textwidth}
    \importlocalfigure{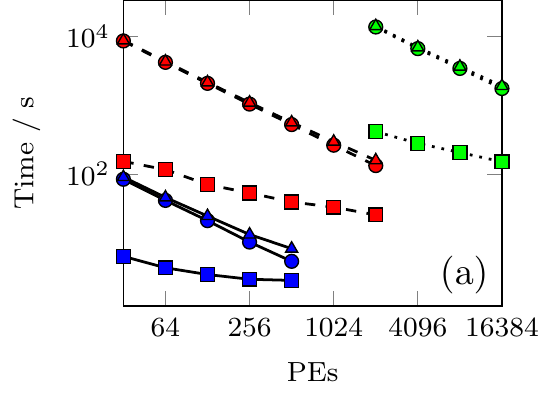}
  \end{minipage}~
  \begin{minipage}{0.33\textwidth}
    \importlocalfigure{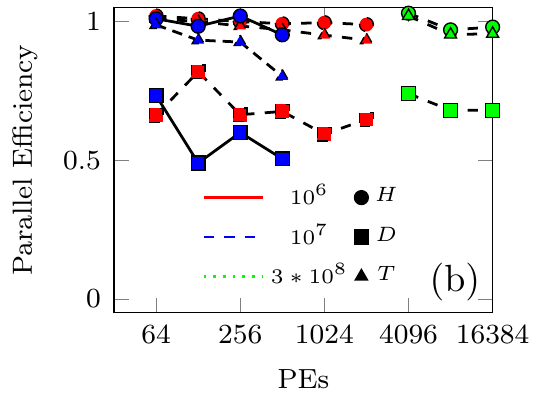}
  \end{minipage}~
  \begin{minipage}{0.33\textwidth}
    \importlocalfigure{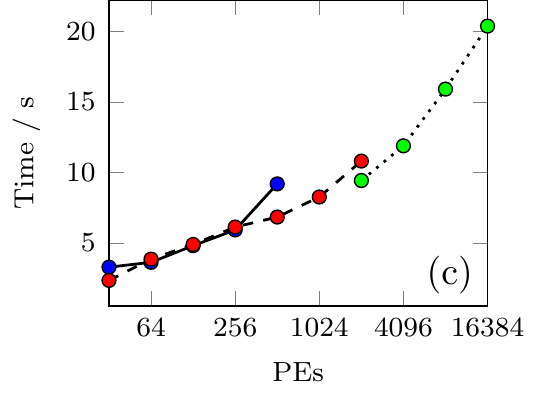}
  \end{minipage}
  \caption{Strong scaling of $N_{tdets}=10^6,10^7$, and $3*10^8$ ASCI. (a) Illustrates the strong scaling of the Hamiltonian generation, Davidson iterations 
  and overall application performance for each system while (b) illustrates the corresponding parallel efficiency. (c) Plots the ratio the the largest to the smallest number of non-zero matrix elements over the PEs, which explains the imbalance in the Davidson iterations.}
  \label{fig:par_ham_eig}
\end{figure}

\Cref{fig:par_ham_eig} illustrates the strong scaling
of the sCI Hamiltonian generation and subsequent
Davidson iterations for Cr$_2$ with $N_{cdets} = 300k$
and $N_{tdets} = 10^6, 10^7$ and $3*10^8$. It is 
clear that the cost of Hamiltonian formation dominates
the variational ASCI calculation at all $N_{tdets}$ considered, and is $O(10x)$ the cost of the Davidson iterations and $O(100x)$ the cost of the ASCI search
discussed in the previous subsection. The Hamiltonian
formation exhibits near linear strong scaling ($>$ 98\%
parallel efficiency), whereas the Davidson iterations
exhibit between 50\%-80\% parallel efficiency. 
The scaling of the former is expected, as each PE performs
nearly the same number of determinant distance
checks in the double loop method (\cref{sec:parallel_eigensolver}) and there is no
communication between PEs.  The scaling of the Davidson
iterations can be understood in terms of \cref{fig:par_ham_eig}(c) where the source of the load imbalance can be attributed to the variance of non-zero
matrix elements among PE. This is due to the fact that the 
ordering of the determinants in the ASCI wave function, as produced by the parallel
ASCI search,
is more or less random, and thus unable to guarantee
regular distributions of non-zero matrix elements. 
This problem is well known in the field of distributed
memory sparse linear algebra \cite{10.1145/2755573.2755595}, as both the local work
and overall communication volume of the parallel SpMV
(\ref{alg:dist_spmv}) are affected by irregular distributions of
non-zero matrix elements. Improving the communication
characteristics of the sCI SpMV will be the subject
of future work by the authors. \Cref{fig:par_ham_eig} also shows the overall parallel efficiency of the ASCI application for the different problem sizes. As the calculation is dominated by the Hamiltonian construction, we see similar parallel efficiencies for the total application, particularly for the larger problems. Overall, the ASCI calculation maintains $>$85\% parallel efficiency for the 
smallest problem ($10^6$) which maintaining $>$95\% parallel efficiency for the largest problem at $3*10^8$ variational determinants out to 16,384 PEs. To the best of the
authors' knowledge, this is the largest variational ASCI calculation to date.

\iffalse
TODO:
\begin{enumerate}
    \item Demonstrate strong / weak scaling of search algorithm viz cdets / tdets. Posit 
    the implementation is embarrassingly parallel.
    \item Demonstrate strong scaling of Hamiltonian construction. Depending on how slow
    the local builds are, discuss implications of faster bit representations and how they
    will not affect overall scaling as there is no communication / minimal load imbalance
    for double loop (number of checks is constant)
    \item Demonstrate scaling for SPMV w/ and w/out METIS - show improvement and discuss
    implications viz cost (prefactor) and scaling trade-offs
    \item Demonstrate scaling of parallel eigensolver (Davidson). In the context of SPMV,
    discuss implications of communication bottlenecks and how they might be improved
    by future advances in communication avoiding Krylov algorithms
    \item Demonstrate for a single case the scaling of the overall calculation. 1B det ASCI?
\end{enumerate}
\fi

\section{Conclusions}
\label{sec:conclusions}

In this work, we have proposed and demonstrated a parallel, distributed memory 
extension for the variational ASCI method. We addressed key concerns for the
dominant steps of the ASCI algorithm, including the development of a parallel
algorithm for the determinant selection, a parallel algorithm and sparse 
storage scheme for the sCI Hamiltonian, and a parallel algorithm for the
variational eigenvalue calculation via the Davidson method. The parallel
search algorithm was enabled by previous work in the context of ASCI-PT2,
where the introduction of determinant constraints offered a conventient and 
robust mechanism to distribute work across PEs. This method was extended to
include higher order determinant constraints than were considered for the ASCI-PT2 method and coupled with a load balancer
to ensure scaling on large compute resources. 
We have demonstrated
the efficacy and performance of the proposed algorithms on a challenging test
system (Cr$_2$) on up to 16,384 cores. The use of 
higher-order determinant constraints yielded a 5-5.5x performance improvement in the ASCI search (triplet constraints) in the strong scaling limit. The overall ASCI calculation was demonstrated to maintain 85\% - 95\% parallel efficiency.
In addition we have demonstrated
stability and continued convergence of the ASCI method into the regime of 
hundred-of-millions of determinants, which to
the best of the authors' knowledge, is the largest
variational ASCI calculation to date. 
%(a couple of larger sCI simulations have been performed previously~\cite{Li2018}).
Although the developments for the parallel search algorithm are ASCI specific,
the developments pertaining to distributed Hamiltonian construction and 
Davidson diagonalization are directly applicable to other sCI methods as well.
These developments indicate a promising future for massively parallel
sCI applications in the years to come.

While these results are promising, there remain open areas for exploration
which will be addressed by the authors in future work. In this work,
the Hamiltonian construction dominates the overall ASCI application by
at least an order of magnitude. Application of more advanced Hamiltonian
matrix element techniques~\cite{asci20_modern} for the local sparse matrix
construction offer high-potential for acheiving further improvements in the future although it is unclear whether such developed will introduct load imbalance. The interplay between efficient Hamiltonian construction and load imbalance warrants further exploration. In addition, the strong 
scaling stagnation of the Davidson iterations can be attributed primarily to
massive load imbalance in the non-zero element distribution across PEs.
In our experience, the existing technologies\cite{metis1,mckee69_reducing} for reordering sparse matrices
to minimize communication volume are insufficient for producing balanced 
partitions for molecular CI calculations. Development of scalable graph reordering 
techniques for the CI problem would be a particular fruitful area of
exploration given massively parallel implementations of sCI.  Lastly, the leverage of bit operations both in the search
and Hamiltonian construction steps of the ASCI algorithm will be particularly
advantageous on current and future GPU architectures and possibly other
hardware accelerators to come. As such, the authors will build upon
initial results\cite{asci20_modern} for GPU applications to the ASCI method
and develop GPU accelerated extensions of the parallel algorithms proposed
in this work.

%In this work we looked at a parallel implementation of the ASCI algorithm.  We
%consider novel approaches to parallel matrix diagonalization, and we explicitly
%implemented a parallel search algorithm and demonstrate strong scaling for our
%approach.  Our approach to parallel search and parallel PT2 is likely to be the
%closest to many other selected CI variants in comparison to a recent parallel
%HCI implementation.  HCI is an algorithm in which the search sum is not
%performed~\cite{holmes2016}, and the PT2 part is performed with a Monte Carlo
%technique~\cite{Li2018}.  Both sums are performed in ASCI with deterministic
%techniques, and the parallel approaches here can be used in other sCI
%approaches in which the sums are calculated in a similar manner.  The main goal
%in working through a parallel approach is to demonstrate that the techniques
%can scale with increasing tasks/processors.  We have put forth algorithms that
%do demonstrate this scaling.  In future works we will look to refine these
%approaches even further and increase the efficiency of the methods.  

\section{Acknowledgment}
DWY and WAdJ were supported from the Center for Scalable Predictive methods for
Excitations and Correlated phenomena (SPEC), which is funded by the U.S.
Department of Energy (DoE), Office of Science, Office of Basic Energy Sciences,
Division of Chemical Sciences, Geosciences and Biosciences as part of the
Computational Chemical Sciences (CCS) program at Lawrence Berkeley National
Laboratory under FWP 12553.  NMT is grateful for support from NASA Ames
Research Center.  CMZ acknowledges financial support from the European Research
Council (ERC), under the European Union’s Horizon 2020 research and innovation
programme, Grant agreement No. 692670 ”FIRSTORM”.  This research used resources
of the National Energy Research Scientific Computing Center (NERSC), a U.S.
Department of Energy Office of Science User Facility located at Lawrence
Berkeley National Laboratory, operated under Contract No. DE-AC02-05CH11231
using NERSC award BES-ERCAP-M3196.  NMT acknowledges funding from the NASA ARMD
Transformational Tools and Technology (TTT) Project.   Some calculations were
performed as part of the XSEDE computational Project No. TG-MCA93S030 on
Bridges-2 at the Pittsburgh supercomputer center.

\bibliography{refs}

\end{document}